%% file: sample-journal.tex
\documentclass[acmlarge]{acmart}
\usepackage{booktabs} 
\usepackage{longtable}
\usepackage[ruled]{algorithm2e} 
\usepackage{todonotes}
\usepackage{xcolor}
\usepackage{subcaption}
\usepackage{printlen}
\usepackage{pbox}

\acmVolume{0}
\acmNumber{0}
\acmArticle{0}
\acmYear{2018}
\acmMonth{4}

\begin{document}
\title{Extraction of Behavioral Features from Smartphone and Wearable Data}

\author{Afsaneh Doryab}
\affiliation{%
  \institution{Carnegie Mellon University}
  \city{Pittsburgh}
  \state{PA}
  \postcode{15213}
  \country{USA}}
\email{adoryab@cs.cmu.edu}
\author{Prerna Chikarsel}
\affiliation{%
  \institution{Carnegie Mellon University}
  \city{Pittsburgh}
  \state{PA}
  \postcode{15213}
  \country{USA}}
\email{prerna@cmu.edu}
\author{Xinwen Liu}
\affiliation{%
  \institution{Carnegie Mellon University}
  \city{Pittsburgh}
  \state{PA}
  \postcode{15213}
  \country{USA}}
\email{xinwenl@andrew.cmu.edu }
\author{Anind K. Dey}
\affiliation{%
  \institution{University of Washington}
  \city{Seattle}
  \state{WA}
  \postcode{}
  \country{USA}}
\email{anind@uw.edu}
\renewcommand\shortauthors{Author et al}

\begin{abstract}
The rich set of sensors in smartphones and wearable devices provides the possibility to passively collect streams of data in the wild. The raw data streams, however, can rarely be directly used in the modeling pipeline. We provide a generic framework that can process raw data streams and extract useful features related to non-verbal human behavior. This framework can be used by researchers in the field who are interested in processing data from smartphones and Wearable devices. 
\end{abstract}

%
%

%
%


\maketitle

\input{samplebody-journals}

\end{document}

%% file: samplebody-journals.tex
\section{Introduction}
Mobile phones and wearable devices are now equipped with powerful sensors that allow us to collect information about a user's context and behaviors, including location, communication, environment, phone usage, physical activity and sleep. Existing research has explored the capability of mobile phones and fitness trackers to continuously track and collect information about the daily behavior of users through their sensing channels and to use this data to analyze the state of physical and mental wellbeing such as sleep duration and quality\cite{min2014toss} and depression( \cite{doryab2014detection,saeb2015mobile,saeb2016relationship}). \\ 
The widespread use of such devices for research in the ubiquitous and mobile computing communities including pervasive and mobile health as well as their common set of sensing channels gives rise to providing a comprehensive and generic framework for data collection and processing that can be shared in the research community. Such framework has the following advantages:
\begin{itemize}
\item It provides a generic tool for mobile data processing that can be easily used or adapted by other researchers.
\item The common set of features provides the possibility to replicate other research results in the field.
\end{itemize}
However, so far, research studies with mobile and wearable devices have provided their own processing method often focusing on few data channels, e.g., processing location data only, phone usage traces only, or location and phone usage only. The extracted features has also been far from a comprehensive set and often without providing reasonable details of the extraction process. Modeling behavior and acquiring knowledge from data, however, requires a multi-modal approach considering and processing data from different sources in order to draw an accurate picture of reality. Besides, multiple data sources help covering missing data from those sources.

In this paper, we present our generic feature extraction framework for mobile timeseries data that has been developed in our research lab and used in numerous research analyses on human behavior modeling. 
We  provide a rich set of features representing behaviors and change in behaviors computed over different temporal slices (77805 features in total) that can be adapted and used for different computational problems in modeling and inferring human behavior. 
The following describes the type of data sources and the feature extraction process.


\section{Data Processing and Feature Extraction}
We use the AWARE framework~\cite{ferreira2015aware}, a generic data collection mobile application with supporting backend and network infrastructure to collect passive behavioral data from smartphones and wearables. AWARE is available in both Android and iOS versions. It enables us to collect data from a variety of sources including accelerometer, barometer, applications, battery, bluetooth, calls and messages, wifi, gyroscope, audio, magnetometer, location, contact list, rotation, screen, light, Fitbit, and keyboard. The AWARE framework stores data collected from phone channels and Fitbit on the device and transmits deidentified data to a secure server over a secure network connection when the device is connected to Wi-Fi. 

In this paper we describe extraction of \textit{seven} feature sets: Bluetooth, Calls, Location, Campus Map, Phone Usage, Steps, and Sleep. The campus Map we describe is specific to studies related to college students, but it can easily be adapted to other places. 
Location and Campus Map features capture users' mobility patterns, Calls features capture communication patterns, Bluetooth features can reflect both mobility and communication patterns, and Steps capture physical activity. 
Together they can indicate daily behavior of users.
%
%
%
%
%
%

The feature extraction approach for each of the seven feature sets is described in the following sections. We extract features over the 4 daily time windows, namely morning (6am-12), afternoon (12-6pm), evening (6pm-12am), and night (12am-6am). We also aggregate features over the week, weekdays, and weekends as well as half and full semester (for college students datasets). We also extract behavioral change counterparts for every feature (see section \ref{bcfeatures}). 
\subsection{Bluetooth Features}\label{bluefeatures} Bluetooth features are calculated from the scanned bluetooth addresses recorded by the Bluetooth sensor in the smartphone. Scanned bluetooth addresses can be clustered into the person's own devices (``self'' -- scanned most often), family/ roommate/ office mate's devices (``related'' -- scanned less often than ``self'' but more often than ``others''), and other people's devices (``others'' -- scanned least often) to help us estimate how many different people the person meets, thereby capturing social activity and collocated communication. Since a person may or may not be living with family or a roommate or be sharing an office, we cluster scanned addresses twice. First, the addresses are clustered into two categories for  ``self'' vs. ``others'' (K = 2 clusters), then into three clusters - ``self'' vs. ``related'' vs. ``others'' (K = 3 clusters), and then chose the model which fit the data best out of the two sets. This process is described below.
\begin{enumerate}
\item{We calculated the number of days each unique bluetooth address is scanned at least once. \\That is, $number\_of\_days_{bti}$. }
\item{We calculated the average frequency of each unique bluetooth address. \\That is, $average\_frequency_{bti} = \frac{total\_count_{bti}}{number\_of\_days_{bti}}$. }
\item{We Z-normalized the $number\_of\_days_{bti}$ and $average\_frequency_{bti}$ in order to give equal weight to both while optimizing score in step 4.}
\item{For each bluetooth address, we computed $Score = number\_of\_days_{bti} + average\_frequency_{bti}$.}
\item{We used K-means clustering to cluster $Score$ from step 4 for all bluetooth addresses using K=2 and K=3.}
\item{Model with K=2 is chosen if sum of squared distances between clustered points and cluster centers is smaller than what we get with K=3. Otherwise we chose model with K=3.}
\item{If model with K=2 is chosen, the cluster with higher scores contained the person's own devices (``self''), while the other cluster contained other people's devices (``others''). If the model with K=3 is chosen, the cluster with highest scores contained the person's own devices (``self''), the cluster with lowest scores contained other people's devices (``others''), and the remaining cluster contained devices of the person's partners, roommates, or officemates (``related'').}
\end{enumerate}
Once the bluetooth addresses scanned are clustered into ``self'' vs. ``others'' or ``self'' vs. ``related'' vs. ``others'', we extract the \textit{number of unique devices}, \textit{number of scans of most and least frequent device}, and \textit{sum, average, and standard deviation of the number of scans of all devices} from all devices (\textit{i.e.}, ignoring clusters), ``self'' devices, ``related'' devices, or ``others'' devices.

\subsection{Calls Features}\label{callsfeatures} Calls features are calculated using the call logs from the smartphone. We extract the following features:

\textit{Number and duration of incoming, outgoing, and missed calls to everyone, family members, friends off-campus, and friends on-campus}, \textit{number of correspondents overall}, and \textit{number of correspondents who are family members, friends off-campus or friends on-campus}.

\subsection{Location Features}\label{locfeatures} Location features are derived from the Location sensor of the smartphone which records the user's GPS coordinates. We extract the following Location features:

\textit{Location variance} (sum of the variance in latitude and longitude coordinates), \textit{log of location variance}, \textit{ total distance traveled}, \textit{average speed}, and \textit{variance in speed}.
%
%
%
%
\textit{Circadian movement} \cite{saeb2015mobile} is calculated using the Lomb-Scargle method \cite{press1989fast}. It encodes the extent to which a person's location patterns follow a 24-hour circadian cycle. Then, we carry out the following processing steps:
\begin{enumerate}
\item Speed of the person is calculated from the distance covered and time elapsed between two samples. Samples with speed $>$ 1 km/h are labeled as ``moving'', else ``static''.
\item Samples labeled as ``static'' are clustered using DBSCAN \cite{ester1996density} to find significant places visited by the person. When we cluster all data and extract each feature per week or per half-semester, we use global clusters. When we first split the data into weeks or half-semesters and then extract features from each temporal slice, we use local clusters. 
\end{enumerate}
These steps allow us to extract: \textit{number of significant places}, \textit{number of transitions between places}, \textit{radius of gyration} \cite{canzian2015trajectories}, \textit{time spent at top-3 (most frequented) local and global clusters}, \textit{percentage of time spent moving}, and \textit{percentage of time spent in insignificant or rarely visited locations} (labeled as -1 by DBSCAN). We also calculate \textit{statistics related to length of stay at clusters} such as maximum, minimum, average, and standard deviation of length of stay at local and global clusters. \textit{Location entropy} and \textit{normalized location entropy} across local and global clusters are also calculated. Location entropy will be higher when time is spent evenly across significant places. Calculating features for both local and global clusters allow us to capture different behaviors related to the user's overall location patterns (global) and the user's location patterns within a time slice (local). For example, time spent at top-3 global and local clusters captures the time spent at places of overall significance to the user and places significant to the user in a particular time slice (\textit{e.g.,} mornings on weekends). 

We assume the place most visited by the person at night to be their home location. To approximate the home location, we perform steps (a) and (b) above on the location coordinates from all nights (12am to 6am) and assume the center of the most frequented cluster to be the person's home location center. Since we don't know the radius of the home, we calculate two home-related features \textit{time spent at home assuming home to be within 10 meters of the home location center}, and \textit{time spent at home assuming home to be within 100 meters of the home location center}. 

\subsection{Places Features}\label{campusfeatures} We also extract the user's location patterns in relation to their surroundings, e.g., a college campus. First, we obtain a map of the persons' surrounding area. Then, we mark out the place boundary and different types of buildings by creating polygons on Google Maps using GmapGIS\footnote{\url{http://www.gmapgis.com/}}. For example, for a campus map, we annotate eight types of buildings and spaces -- Greek houses that hold the most social events, all Greek houses, student apartments, residential halls, athletic facilities, green spaces, academic, and off-campus). Then, we extract the following features for each type of space: \textit{time spent at each location type in minutes}, \textit{percentage time spent at each location type}, \textit{number of transitions between different spaces}, \textit{number of bouts (or continuous periods of time) at space}, \textit{number of bouts during which person spends 10, 20, or 30 minutes at the same space}, and \textit{minimum, maximum, average, and standard deviation of length of bouts at each space}. 
%
%
%
%
%
%
%
%

Campus map features also include two multimodal features -- \textit{study duration} and \textit{social duration}. These features fuse data from Location, Phone Usage, Conversation, and Steps sensors. Study duration is calculated by fusing location type labels with data from the phone usage and steps sensors. A person is assumed to be studying if they spent 30 minutes or more in an academic building while being sedentary (fewer than 10 steps) and having no interaction with their phone. Social duration is calculated by fusing location type labels with data from the conversation sensor. A person is assumed to be social if they spent 20 minutes or more in any of the residential buildings or green spaces and the conversation sensor inferred human voice or noise for 80\% or more of that time.

%
%
%
%

\subsection{Phone Usage Features}\label{phoneusagefeats} Phone Usage features are calculated using the screen status sensor in the smartphone, which recorded screen status (on, off, lock, unlock) over time. We extract the following phone usage features:

\textit{Number of unlocks per minute}, \textit{total time spent interacting with the phone}, \textit{total time the screen is unlocked}, \textit{the hour of the days the screen is first unlocked or first turned on}, \textit{the hour of the days the screen is last unlocked, locked, and turned on}, and \textit{the maximum, minimum, average, and standard deviation of length of bouts (or continuous periods of time) during which the person is interacting with the phone and when the screen is unlocked}. A person is said to be ``interacting'' with the phone between when the screen status is ``unlock'' and when the screen status is ``off'' or ``lock''. 

\subsection{Sleep Features}\label{slpfeatures} Sleep features are calculated from the sleep inferences (asleep, restless, awake, unknown) over time returned by the Fitbit API. The following features are calculated:

\textit{Number of asleep samples}, \textit{number of restless samples}, \textit{number of awake samples}, \textit{number of unknown samples} (still detected as sleep), \textit{weak sleep efficiency} (sum of number of asleep and restless samples divided by sum of number of asleep, restless, and awake samples), \textit{strong sleep efficiency} (sum of number of asleep samples divided by sum of number of asleep, restless, and awake samples), \textit{count, sum, average, maximum, and minimum length of bouts during which the person is asleep, restless, or awake} as well as the \textit{start and end time of longest and shortest bouts during which the person is asleep, restless, or awake}. 

\subsection{Steps Features}\label{stepsfeatures} Steps features are calculated from the step counts over time returned by the Fitbit API. The following features are calculated:

\textit{Total number of steps} and  \textit{maximum number of steps taken in any 5 minute period} are extracted as features. Other features are extracted from ``bouts'', where a ``bout'' is a continuous period of time during which a certain characteristic is exhibited. Examples of such features include \textit{total number of active or sedentary bouts} \cite{bae2016using}, and \textit{maximum, minimum, and average length of active or sedentary bouts}. We also calculated \textit{minimum, maximum, and average number of steps over all active bouts}. A bout is said to be sedentary if the user takes less than 10 steps during each 5 minute interval within the bout. As soon as the user takes more than 10 steps in any 5 minute interval, they switch to an active bout.

\subsection{Behavioral Change Features} \label{bcfeatures}
Behavioral change features capture changes in behaviors over a certain period of time, e.g, $n$ weeks. These features can be abstractly characterized as the change in slope for each behavioral feature over that period. For this purpose, we only use features computed weekly (\textit{i.e.}, using granularity ``weeks''). 
We compute the behavioral change feature for each behavioral feature using their weekly values over the $n$ week and a method similar to \cite{wang2015smartgpa}: 
\begin{itemize}
\item \textit{Slope:} We fit a linear regression model to the values of the feature over $n$ weeks. ``Slope'' is the slope of this linear regression line. 
\item \textit{Slope first half and second half:} We fit two separate linear regression models to the values of the feature over weeks $1 to m$ and weeks $m to n$. ``Slope first half'' and ``Slope second half'' are the slopes of these linear regression lines.
\item \textit{Breakpoint:} Each breakpoint is the week after which the behavior (represented by the feature value) begins to change. This is calculated by fitting a piecewise linear regression model with two segments with each of the $n$ weeks as a breakpoint. ``Breakpoint'' is the week that when used as a breakpoint gives the best model as determined by Bayesian Information Criterion (BIC). 
\item \textit{Slope before and after Breakpoint:} A piecewise linear regression model with two segments is fit to the feature values over $n$ weeks with the final ``Breakpoint''. The slope of the first line segment is ``Slope before Breakpoint'' and the slope of the second line segment is the ``Slope after Breakpoint''.
\end{itemize}

\section{SUMMARY}
In this paper, we described our method in extracting numerous behavioral features from data collected from smartphones and wearable devices. The code is available upon request. We plan to further extend this component to include more features from behavioral and physiological signals.

\bibliographystyle{ACM-Reference-Format}
\bibliography{sample-bibliography}